\documentclass{moriond}
\usepackage{float}
\usepackage{amsmath}
\usepackage[caption=false]{subfig}
\def\Journal#1#2#3#4{{#1} {\bf #2}, #3 (#4)}

\def\NCA{\em Nuovo Cimento}
\def\NIM{\em Nucl. Instrum. Methods}
\def\NIMA{{\em Nucl. Instrum. Methods} A}
\def\NPB{{\em Nucl. Phys.} B}
\def\PLB{{\em Phys. Lett.}  B}
\def\PRL{\em Phys. Rev. Lett.}
\def\PRD{{\em Phys. Rev.} D}
\def\ZPC{{\em Z. Phys.} C}

\def\st{\scriptstyle}
\def\sst{\scriptscriptstyle}
\def\mco{\multicolumn}
\def\epp{\epsilon^{\prime}}
\def\vep{\varepsilon}
\def\ra{\rightarrow}
\def\ppg{\pi^+\pi^-\gamma}
\def\vp{{\bf p}}
\def\ko{K^0}
\def\kb{\bar{K^0}}
\def\al{\alpha}
\def\ab{\bar{\alpha}}
\def\be{\begin{equation}}
\def\ee{\end{equation}}
\def\bea{\begin{eqnarray}}
\def\eea{\end{eqnarray}}
\def\CPbar{\hbox{{\rm CP}\hskip-1.80em{/}}}
\newcommand{\MM}[1]{{\bf \color{red} #1}}
\def\msun{\mathrel{{M_{\odot} }}}

\newcommand{\Photo}{\includegraphics[height=35mm]{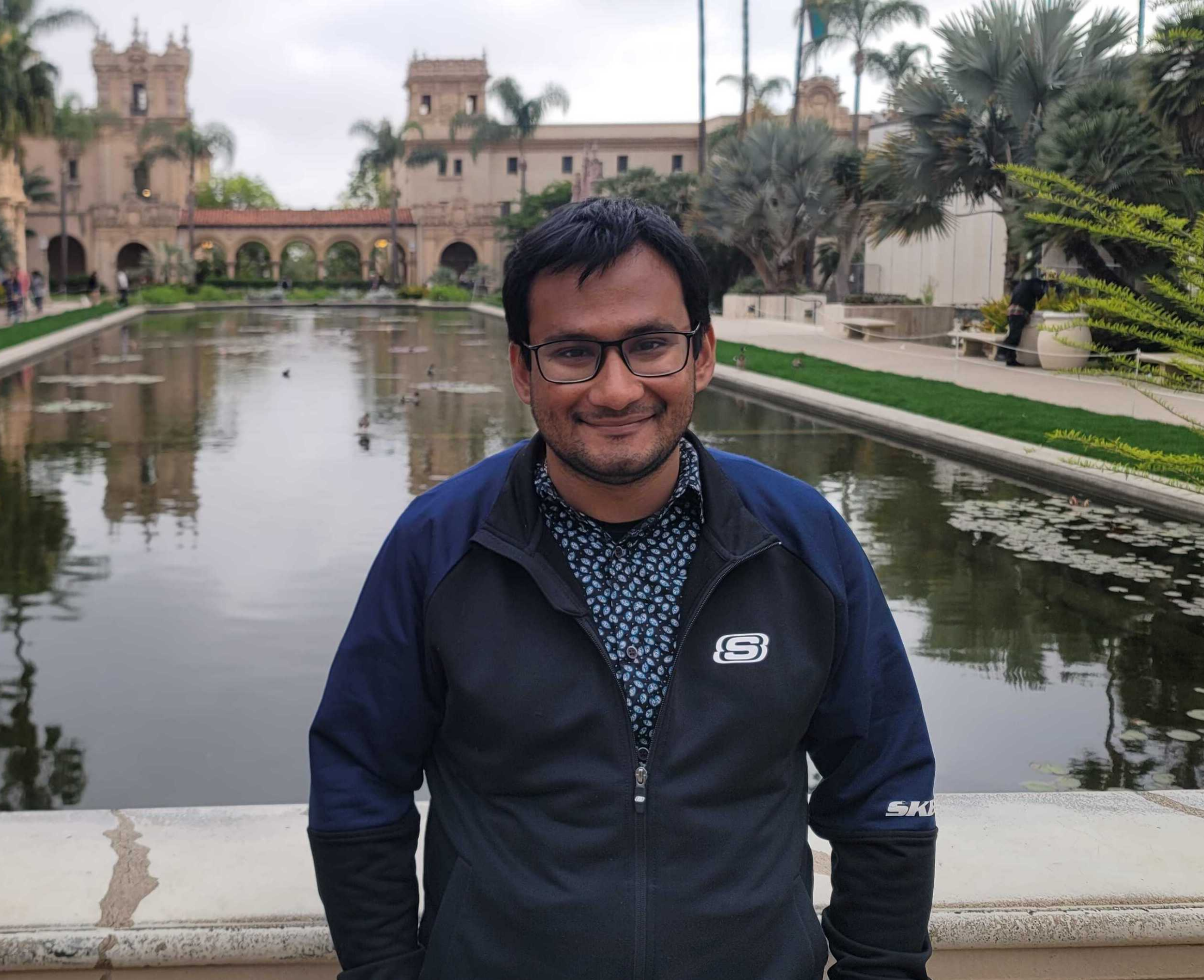}}

\begin{document}
\vspace*{4cm}
\title{SEARCHING FOR SUPERNOVA NEUTRINOS WITH GW MEMORY TRIGGERS}

\author{ MAINAK MUKHOPADHYAY }

\address{Department of Physics, Arizona State University,\\
Tempe, AZ 85287-1504, USA.}

\maketitle\abstracts{
Anisotropic neutrino emission from a core-collapse supernova (CCSN) causes a permanent change in the local space-time metric, called the gravitational wave (GW) memory. Long considered unobservable, this effect will be detectable in the near future, at deci-Hertz GW interferometers. I will present a novel idea, where observations of the neutrino GW memory from CCSNe will enable time-triggered searches of supernova neutrinos at megaton (Mt) scale detectors, which will open a new avenue to studying supernova neutrinos.
}

\section{Introduction}
\label{sec:intro}

In the current era of multi-messenger astronomy, GW, neutrino, photon, and cosmic ray observations are combined to extract information about astrophysical sources and phenomena in the universe~\cite{Engel:2022yig}. Core-collapse supernovae (CCSNe) have been one of the prime focuses of multi-messenger astronomy~\cite{Nakamura:2016kkl}. Amongst the messengers from CCSNe, neutrinos are the most dominant ones. Around, $3 \times 10^{53}$ ergs ($\sim 99\%$) of the gravitational binding energy of a star is emitted in neutrinos over a time scale of $\sim 10$ s. But, the majority of CCSNe have been observed through optical telescopes and we have observed supernova neutrinos from only one CCSN, SN1987A~\cite{Burrows:1987zz}. This makes the aspect of detecting supernova neutrinos one of profound interests and it leads to an obvious question - \emph{why is it that we have supernova neutrino observations from just one CCSN?}

\begin{figure*}
\begin{center}
\subfloat[(a)]{\includegraphics[width=0.49\textwidth]{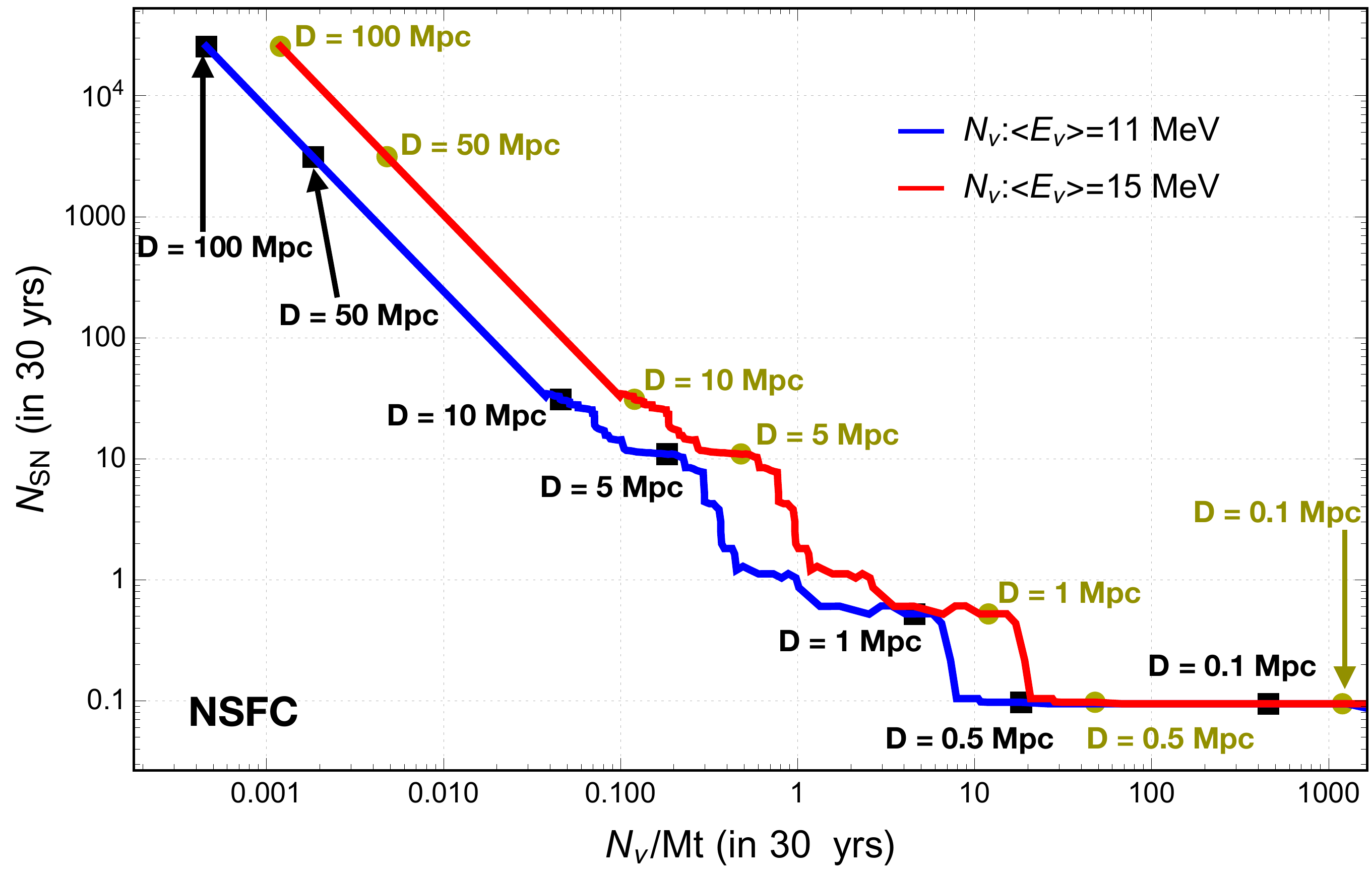}\label{fig:nsn_nnu}} \hfill
\subfloat[(b)]{\includegraphics[width=0.49\textwidth]{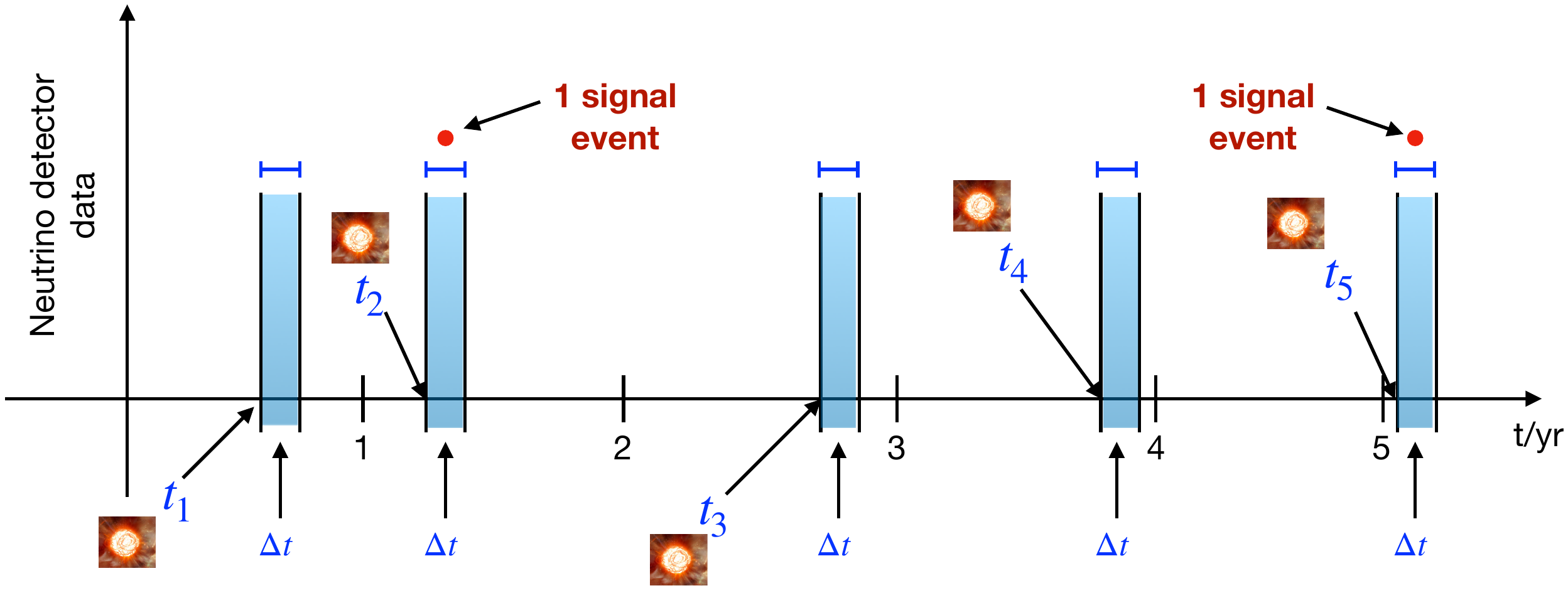}\label{fig:schematic}} \hfill
\end{center}
\caption{(a) The total number of CCSNe versus the total number of neutrinos obtained per megaton from those supernovae in 30 years (see text for details).  (b) Schematic diagram to illustrate the idea of memory-triggered supernova neutrino detection.
}
\end{figure*}

The answer to this can be seen in Fig.~\ref{fig:nsn_nnu}. We have plotted the total number of core-collapse supernovae in 30 years versus the total number of neutrinos that could be obtained per megaton from the supernovae in 30 years. This is done for mean supernova neutrino energy of $11$ and $15$ MeV. The distance from earth is marked by markers in both cases. We assume successful supernovae, that is, neutron-star forming collapses (NSFCs) only. The details of the parameters and the formulae required to obtain the plot will be discussed later. But the evident fact is for nearby distances which would lead to very high statistics, the rate of core-collapse supernova is extremely low. For large distances, the number of core-collapse supernovae is very large but because the neutrino flux falls off as $1/D^2$, the statistics is extremely low.

The current situation on observing CCSN neutrinos is - with the upcoming neutrino detectors, we will manage to have neutrinos observed from nearby CCSNe ($D\leq 3$ Mpc) with significant statistics~\cite{PhysRevLett.95.171101}. The diffuse supernova neutrino background (DSNB) neutrinos will provide information from CCSNe from $\mathcal{O}(1)$ Gpc distances and beyond~\cite{Beacom:2010kk,2017hsn..book.1637L}. This leaves neutrino observations from CCSNe in the local universe ($D \sim 3 - 100$ Mpc) to be an area requiring considerable focus. Ideas proposing multi-megaton neutrino detectors~\cite{Kistler:2008us}, using GW detectors as triggers~\cite{PhysRevLett.103.031102} exist, but are limited to a few Mpc. In this work, we present a summary of a novel idea of using the neutrino GW memory effect to perform time-triggered searches for supernova neutrinos~\cite{Mukhopadhyay:2021gox}, which would in essence be sensitive to $\mathcal{O}(100)$ Mpc.

A schematic representation of the main idea is shown in Fig.~\ref{fig:schematic}. The core-collapse supernova leads to an observable GW memory signal in the GW detectors say at different times, $t_1, t_2, t_3, t_4, \dots$. We know the time duration of a supernova neutrino burst, $\Delta t \sim 10$ s. Now, one can go back to the neutrino detector data and analyze the events within time-windows, $t_1 + \Delta t, t_2 + \Delta t, t_3 + \Delta t, t_4 + \Delta t, \dots$. Since, the background in each of the time-windows is very low, owing to the time duration of just $\sim 10$ s; the probability of finding a signal event (in this case a supernova neutrino) is high. This method of reducing the background events by using a triggered time-window of a short duration allows collecting a clean sample of signal events over time.

In Sec.~\ref{sec:ccsn} we briefly discuss the dynamics of a CCSN and the main neutrino emission phases. Sec.~\ref{sec:gwm} focuses on the neutrino GW memory effect from a CCSN. In Sec.~\ref{sec:formalism}, we discuss the formalism to calculate the number of memory-triggered neutrino events possible. We discuss our main results in Sec.~\ref{sec:results} and conclude in Sec.~\ref{sec:conclusion}.

\section{Core-collapse supernova neutrinos}
\label{sec:ccsn}

This section provides a very short review on the dynamics and neutrino emission from CCSN~\cite{Giunti:2007ry,Janka:2006fh}. A CCSN is the death of a massive star ($M \geq 8 \msun$). The massive stars go through advanced stages of nuclear burning until iron starts forming in the core. As iron forms in the core, it triggers a set of events which finally leads to a pressure loss followed by the collapse of the stellar core. The inner parts of the core reach nuclear densities ($\sim 10^{13}$g/cm$^3$), at which matter has been shown to be incompressible. The infalling matter from the outer layers of the core, stops abruptly because of this and produces a forward moving shockwave. The shockwave propagates outwards but it keeps accreting matter, leading it to stop eventually for a fraction of a second. Following this, it is either re-energized when neutrinos deposit their energy resulting in a successful supernova explosion, leading to the formation of a neutron star (hence also called neutron star forming collapses (NSFCs)) or, sometimes the shock just dies down resulting in a failed supernova, leading to a black hole being formed (hence also called black hole forming collapses (BHFCs)). The main neutrino emission phases from a CCSN are:
\begin{itemize}
\item \textbf{Neutronization burst phase:} The sharp peak in the neutrino luminosity right after the core bounce is known as the neutronization burst. This phase is mostly dominated by electron neutrinos ($\nu_e$) and lasts for $\sim 2$ ms post-bounce.

\item \textbf{Accretion phase:} The accretion phase lasts from $\sim 0.1 - 1$ s post-bounce. During this phase, infalling matter keeps accreting on the forward moving shock. The neutrino emission is nearly thermal and of the order $\sim 10^{52}$ ergs. Radial instabilities develop in this phases because of turbulence, hydrodynamics instabilities and standing accretion shock instability (SASI). Ths leads the neutrino emission in this phase being anisotropic, as has been shown by numerical simulations~\cite{kotake2009ray,Vartanyan:2020nmt}.

\item \textbf{Cooling phase:} In case of a NSFC, the proto-neutron star formed, cools by emitting thermal neutrinos of all flavors characterizing the cooling phase. This phase lasts $\sim 10$ s. Not much is known about the anisotropy in neutrino emission in this phase.
\end{itemize}

\section{The neutrino gravitational wave memory effect}
\label{sec:gwm}

The gravitational wave memory effect is the \emph{permanent} distortion of the local space-time metric caused due to the propagation of a GW. At the linear level, it is caused by gravitationally unbound systems that have anisotropic emission of energy in the form of mass or radiation. Unfortunately, although theoretically this phenomenon has been long known, we are yet to observe it. This is because, it is a sub-dominant effect thus requires a very powerful source, some anisotropy\footnote{Isotropic emissions will lead to the quadrupole moment to be zero, resulting in no GW emission.}, and detectors that would probe the frequency regimes of interest.

CCSNe have been considered to be an ideal source for GW memory observations. The energy emitted in neutrinos is $\sim 10^{53}$ ergs, as discussed in Sec.~\ref{sec:intro}, there is some anisotropy ($\sim 10^{-3} - 10^{-2}$) in the neutrino emission and the frequency regimes of interest to us is the low frequency (deci-Hz) scale\footnote{Since the neutrino emission time scale is $\mathcal{O} (10)$ s, the frequency is $\sim 0.01$ Hz.} which will be explored by a bunch of planned next-generation GW detectors like DECIGO~\cite{2021PTEP.2021eA105K}, BBO~\cite{PhysRevD.83.044011}. Recently, the neutrino GW memory effect was explored~\cite{Mukhopadhyay:2021zbt}, in the context of building a new phenomenological description, highlighting its detectability, and physics potential in the present context. 

\begin{figure*}
\begin{center}
\includegraphics[width=0.48\textwidth]{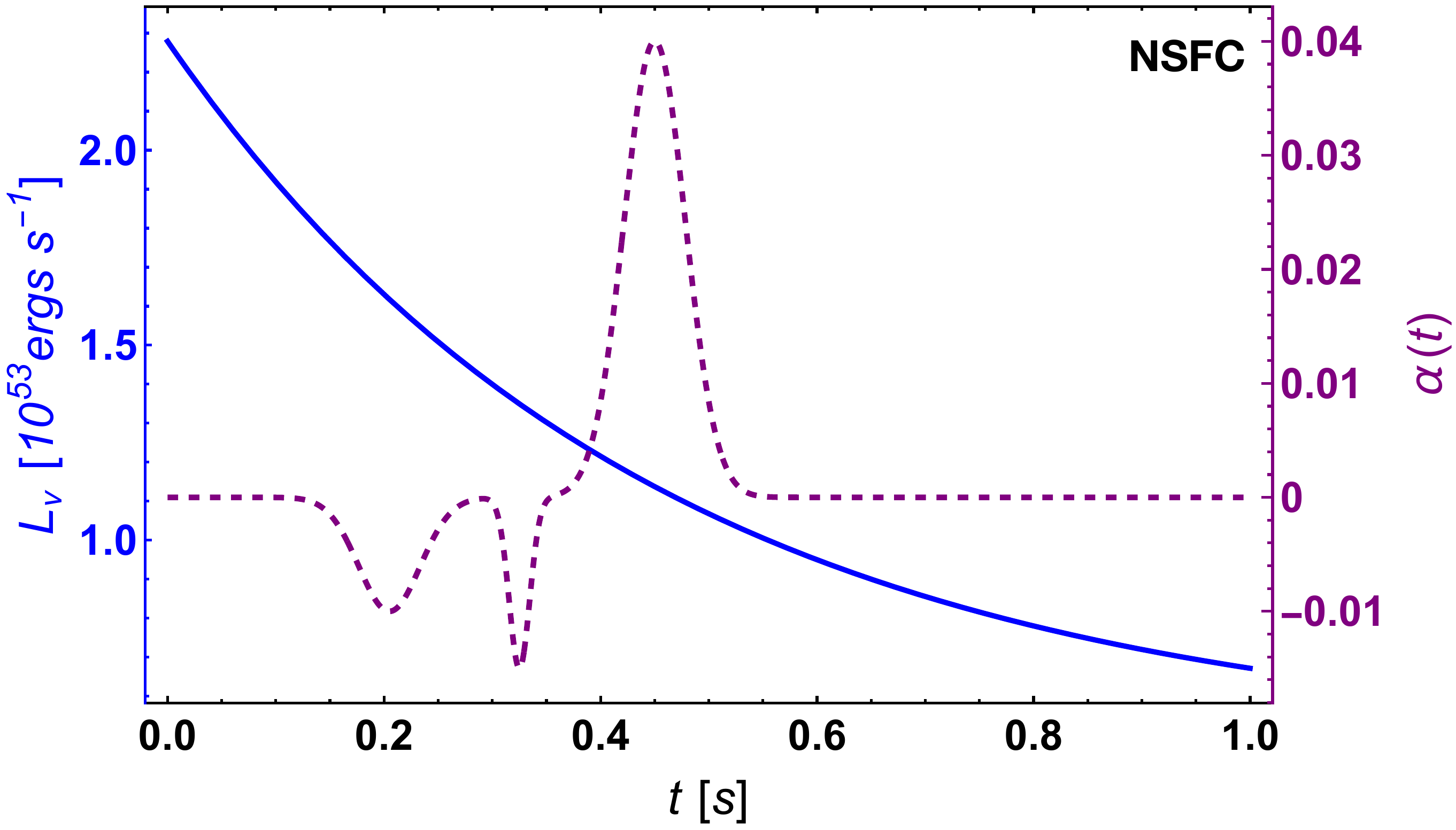} \hfill
\includegraphics[width=0.48\textwidth]{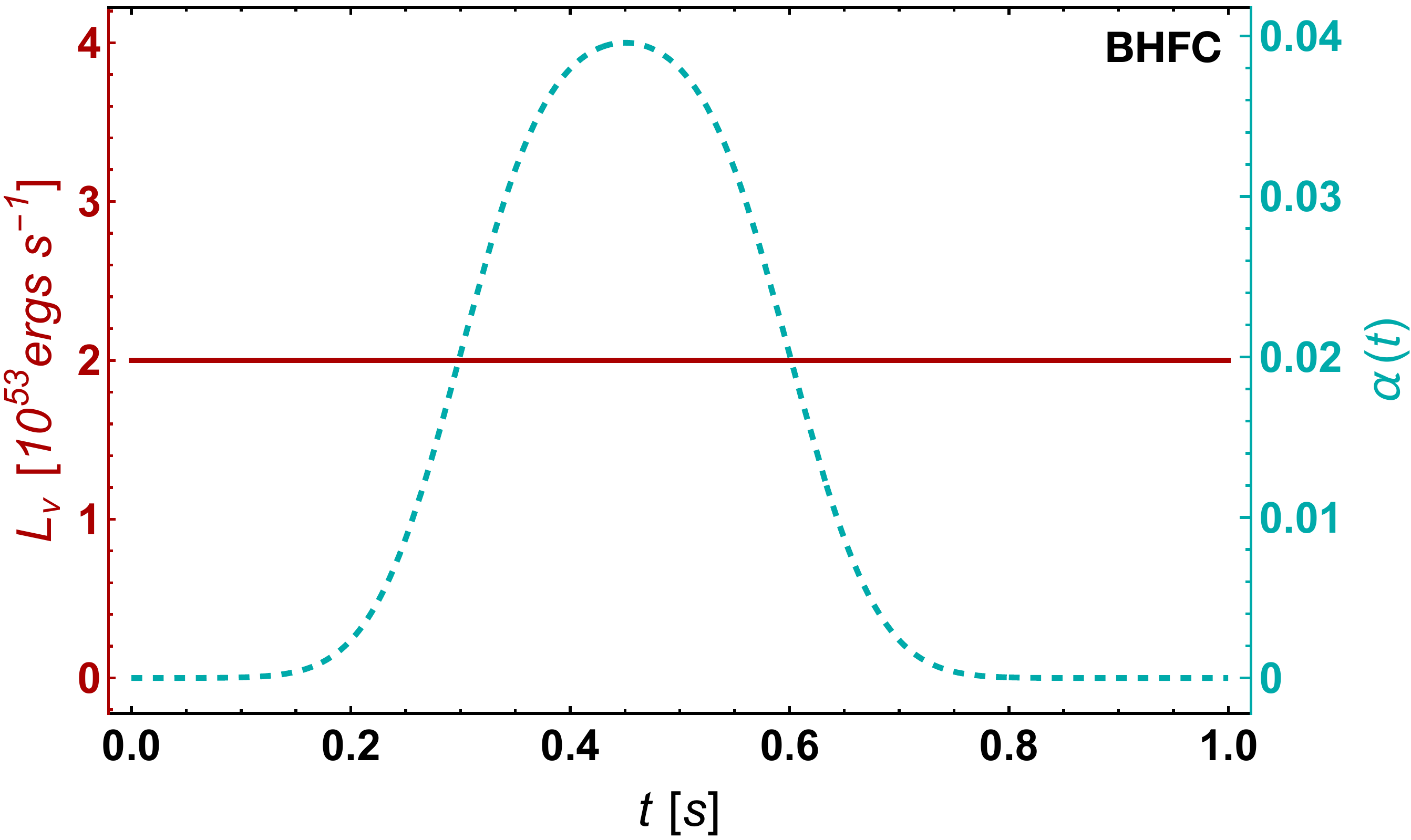} \hfill
\caption{
The neutrino luminosity $L_{\nu} (t)$ and anisotropy parameter $\alpha (t)$ for NSFC (right) and BHFC (left).  
}
\label{fig:models}
\end{center}
\end{figure*}

The neutrino GW memory strain is given by~\cite{1997A&A...317..140M,Mukhopadhyay:2021zbt},
\be
\label{eq:strain}
h(r,t) = \frac{2G}{r c^4} \int_{-\infty}^{t-r/c} dt^{\prime} L_{\nu}(t^{\prime}) \alpha(t^{\prime})\,.
\ee
where $c$ is the speed of light, $t$ is the time post bounce and $G$ is the universal gravitational constant. $L_\nu (t)$ is the all-flavors neutrino luminosity and $\alpha (t)$ is the  anisotropy parameter. It is evident from Eq.~\ref{eq:strain}, that the GW strain depends on time-integral of the product of the neutrino luminosity and the anisotropy parameter. We consider two distinct scenarios and model the neutrino luminosity and anisotropy accordingly (shown in Fig.~\ref{fig:models}). In particular, the NSFC is based on the Ac3G model and the BHFC is based on the LAc3G model~\cite{Mukhopadhyay:2021zbt}. In both cases we consider a non-zero anisotropy only in the accretion phase. For the NSFC, we model the neutrino luminosity in the accretion phase using a decaying exponential. The anisotropy parameter consists of three-Gaussian bumps. In case of the BHFC, the neutrino luminosity in the accretion phase is modeled using a constant. The anisotropy parameter is broader, that is, it lasts for a longer duration and is modeled using three-Gaussians combined together.


\section{Formalism}
\label{sec:formalism}


The formalism mainly relies on, the probability that the GW wave detector sees the memory signal from a CCSN in some galaxy that is at a distance say $r$ from the earth, the number of supernova neutrinos that would be detectable from the same CCSN in a neutrino detector and the number of such CCSN possible at that distance, which will depend on the CCSN rate at that distance.
%
%
The probability that the GW detector detects the memory signal depends on the signal-to-noise ratio (SNR) - which depends on the signal strength, distance to the source, and characteristics (power spectral noise density) of the detector. One can then define the probability of detection $P_{det}^{GW}$ given a fixed false alarm probability~\cite{Jaranowski:1999pd}. The probability of detection depends on the SNR, which in turn as mentioned depends on the distance to the source. 
%

For the neutrino detectors, we restrict ourselves to water Cherenkov detectors (like HyperK~\cite{Abe:2011ts}). The main channel of detection considered is inverse-beta decay (IBD)\footnote{IBD: $\bar{\nu}_e + p \rightarrow e^+ + n$.}. The time-integrated $\bar{\nu}_e$ flux $\Phi (r, E_\nu)$ is given by the well-known analytical quasi-thermal spectra~\cite{Keil:2002in}, where we vary the time-integrated average energy for the electron antineutrino flavor between~\cite{Kresse:2020nto} $11 - 15$ MeV for NSFC, and $15 - 20$ MeV for BHFC.
The predicted number of  events in the detector from a CCSN at distance $r$ is:
\be
N (r) = \int_{E^{th}_\nu}^{E^{max}_\nu} N_{p}  \eta  \sigma(E_\nu)  \Phi(r, E_\nu)\ dE_\nu \,,
\label{eq:nev}
\ee
where $N_{p}$ is the number of target protons, $\eta$ is the detection  efficiency, and $\sigma (E_\nu)$ is the IBD cross-section. We take an energy interval $\left[ E^{th}_\nu, E^{max}_\nu\right]=[19.3,50]\ \text{MeV}$ to avoid the spallation background at low energy and the atmospheric neutrino background at high energy~\cite{Abe:2011ts}. 
%

Below $11$ Mpc the rate of CCSNe is taken from the CCSNe rates in the individual galaxies~\cite{Nakamura:2016kkl}. Beyond $11$ Mpc, we use a constant volumetric rate of CCSN given by\footnote{We ignore the redshift $z$ for distances concerning us.}, $R_{SN} = 1.5 \times 10^{-4}\ \text{Mpc}^{-3}\text{yr}^{-1}$. The cumulative rate of CCSN can be calculated from this analytically by performing a suitable volume integral. 

\section{Memory-triggered supernova neutrino search: Results}
\label{sec:results}

\begin{figure*}
\begin{center}
\includegraphics[width=0.48\textwidth]{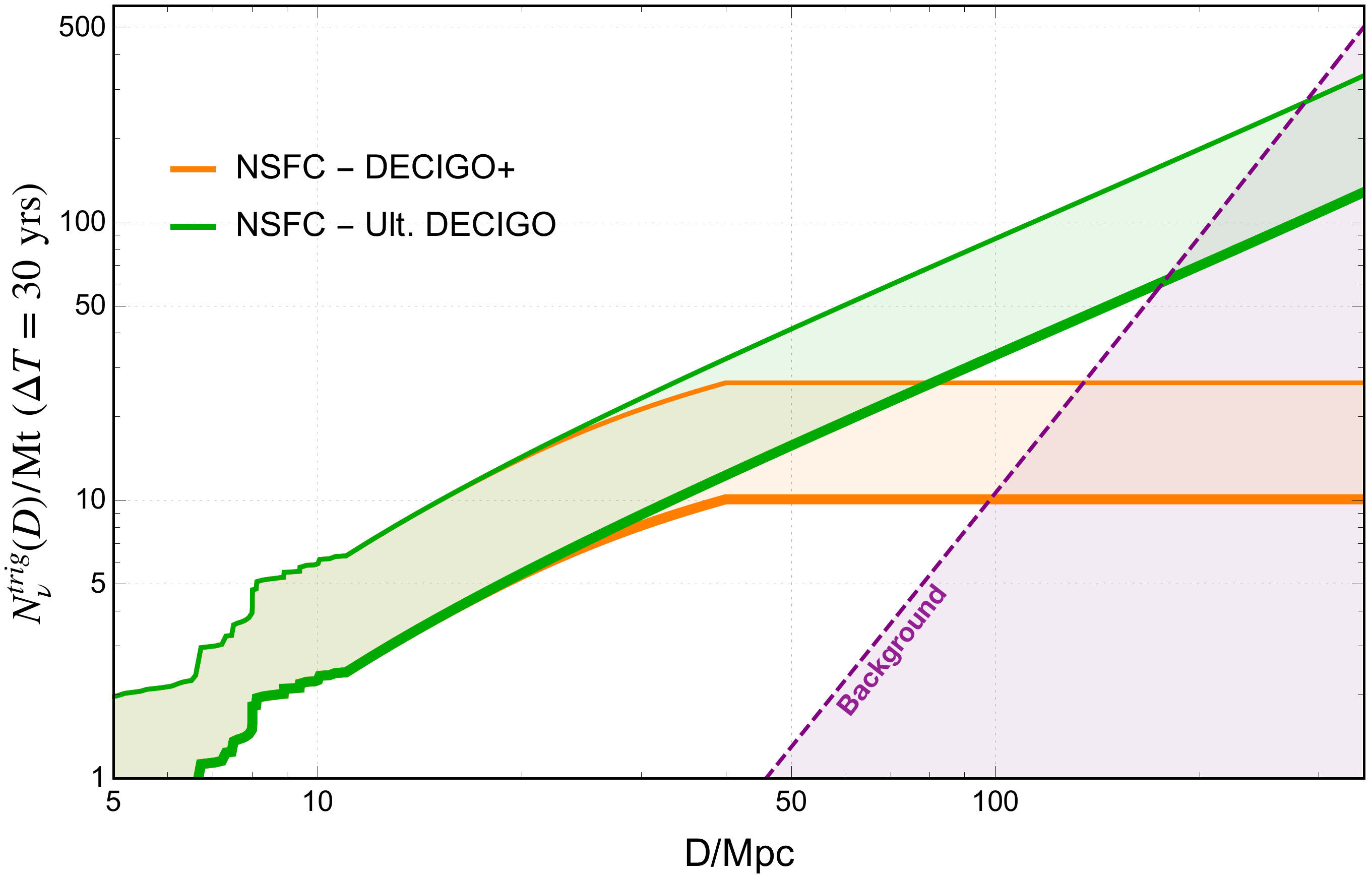} \hfill
\includegraphics[width=0.48\textwidth]{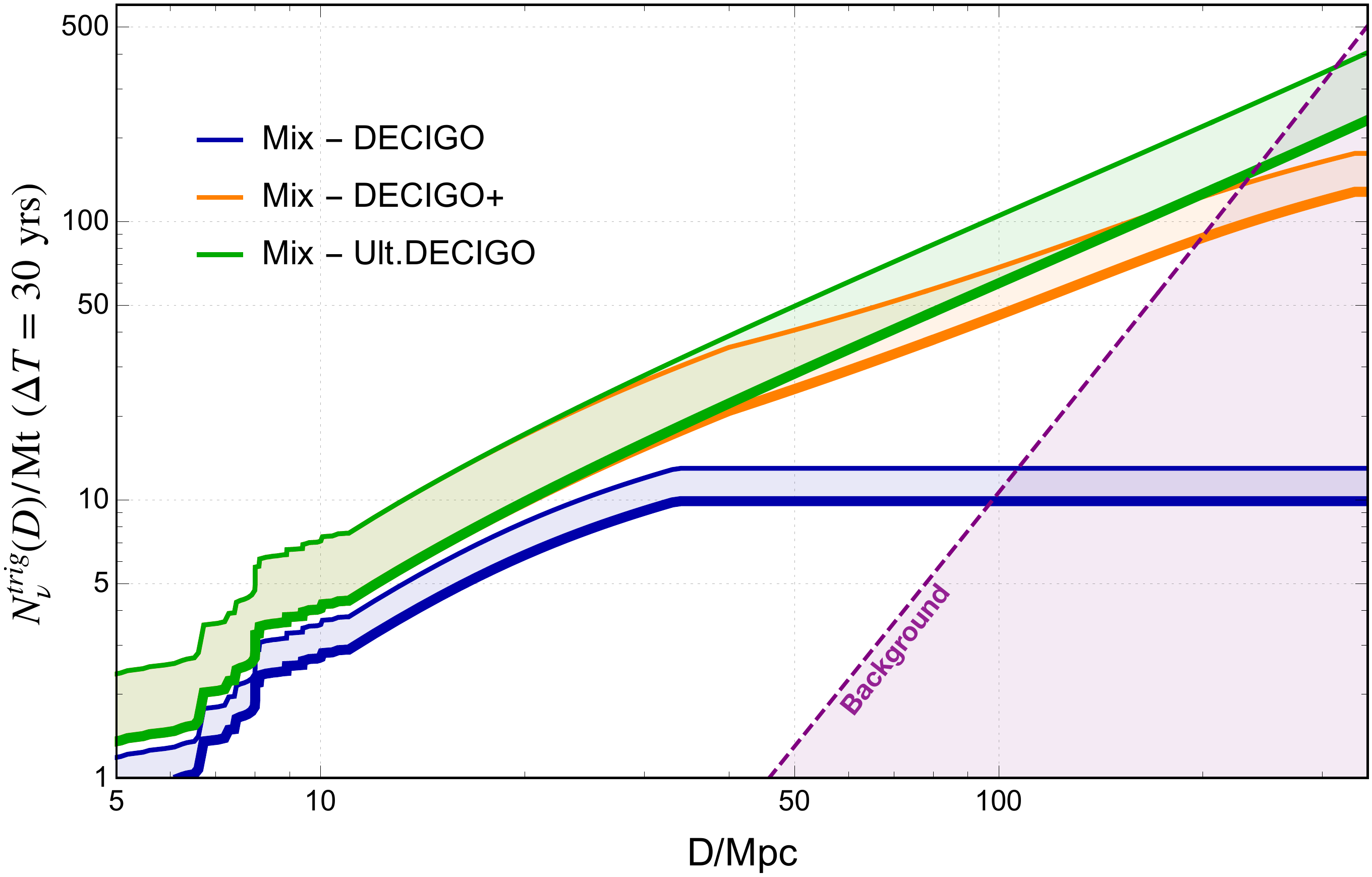} \hfill
\caption{
Number of memory-triggered neutrino events from CCSN at distance $r<D$, as a function of $D$, for a  Mt water Cherenkov detector and $\Delta T = 30$ years. The triggered-background is also shown.
}
\label{fig:nnu_d}
\end{center}
\end{figure*}

The total number of memory-triggered neutrino events from all CCSN within a distance $D$ is given by,
\be
\label{eq:trigevents}
    N^{trig}_{\nu}(D)=\Delta T \sum_{j,r_j<D} R_j N(r_j)P_{det}^{GW}(r_j)\,,
\ee
where, $\Delta T$ is the detector running time, $R_j$ indicates the CCSN rate in the galaxy $j$. The other elements in the above formula are discussed in Sec.~\ref{sec:formalism}. We show the main results ($ N^{trig}_{\nu}(D)$ as a function of $D$ from Eq.~\ref{eq:trigevents}) in Fig.~\ref{fig:nnu_d}, where we have considered - a) a population composed of just NSFCs, and b) a \emph{mixed} population consisting of $60\%$ NSFCs + $40\%$ BHFCs. The shadings indicate the mean neutrino energy being varied for both the NSFCs and BHFCs. The summation in Eq.~\ref{eq:trigevents} becomes an integral, beyond $11$ Mpc when the analytical rates for CCSN need to used.
For the NSFC population (a), time triggers from DECIGO+ will result in  $N^{trig}_{\nu}\sim 10-30$. For Ult. DECIGO, $N^{trig}_{\nu}\sim 100-300$ is expected. For the mixed population (b), the prospects increase substantially for DECIGO+, where it can help detect $\sim 100$ neutrino events due to its increased distance sensitivity for memory in the BHFC scenario. For Ult. DECIGO the results do not change significantly from the previous case. Even DECIGO in its currently planned state, would help trigger and detect $\sim 10$ supernova neutrinos. We also show the background from the triggered-searches in Fig.~\ref{fig:nnu_d} (shaded violet area). This background associated with a triggered-search is, $N^{trig}_{bckg}(D)=N^{trig}_{SN}(D) \lambda \Delta t$, where $\lambda \simeq 1313$ events/year is the background rate in the detector \cite{Abe:2011ts}, $N^{trig}_{SN}(D)=\Delta T \sum_{j,r_j<D} R_j P_{det}^{GW}(r_j)$ is the number of CCSN memory signals observed. It is evident right away, that the triggered-search background is orders of magnitude lesser than the untriggered background, thus highlighting the efficiency of triggered-searches. Furthermore, we see that even the triggered-background dominates the signal events beyond $\sim 350$ Mpc and hence we restrict our analysis to $4<D<350$ Mpc.  
%
%
\section{Conclusion and Discussions}
\label{sec:conclusion}

The technique described in this work is a new multi-messenger approach, in which, the observation of the neutrino GW memory signals enable time-triggered searches for supernova neutrinos. It could be realized in a few decades with the upcoming deci-Hz GW interferometers and Mt-scale neutrino detectors. This will help us collect samples of supernova neutrinos from the local universe ($\sim 10 - 100$ Mpc) which when compared with galactic SN and DSNB neutrinos may help us perform population averaged energy and luminosity studies, understand the distribution of SN populations including NSFCs and BHFCs, and assist in the joint analysis of CCSNe using EM signals, taking our prospects of analyzing CCSNe through their various messengers a step forward. 
\section*{Acknowledgments}
We thank Zidu Lin, Carlos Cardona and Cecilia Lunardini in collaboration with whom this work was performed. We are grateful to Cecilia Lunardini for comments and careful reading of the manuscript. MM acknowledges funding from the National Science Foundation grant number PHY-2012195 and the Fermi National Accelerator Laboratory (Fermilab) Award No. AWD00035045 during this work.

%
%
\section*{References}

\bibliography{refs}

\end{document}